\documentclass[aoas,nameyear,dvips]{arximspdf}
\usepackage{subenv}
\usepackage{dcolumn}

\doi{10.1214/09-AOAS313}
\volume{4}
\issue{2}
\pubyear{2010}
\firstpage{567}
\lastpage{588}

\makeatletter
\newcolumntype{d}[1]{D{.}{.}{#1}}
\makeatother

\begin{document}
\begin{frontmatter}

\title{Maximum likelihood estimation for social network~dynamics}
\runtitle{Social network dynamics}

\begin{aug}
\author[a]{\fnms{Tom A. B.} \snm{Snijders}\thanksref{t1}\ead[label=e1]{tom.snijders@nuffield.ox.ac.uk}\corref{}},
\author[a]{\fnms{Johan} \snm{Koskinen}\thanksref{t1}\ead[label=e2]{johan.koskinen@nuffield.ox.ac.uk}}\\ \and
\author[b]{\fnms{Michael} \snm{Schweinberger}\thanksref{t2}\ead[label=e3]{michael.schweinberger@stat.psu.edu}}
\runauthor{T. A. B. Snijders, J. Koskinen and M. Schweinberger}
\affiliation{University of Oxford, University of Groningen, University of
Oxford\\
and Penn State University}
\thankstext{t1}{Supported in part by the US National Institutes of Health
(NIH 1R01HD052887-01A2).}
\thankstext{t2}{Supported in part by the Netherlands
Organization for Scientific Research (NWO 401-01-552 and 446-06-029).}

\address[a]{T. A. B. Snijders\\ J. Koskinen\\ Nuffield College \\ New
Road, Oxford OX1 1NF \\ United Kingdom \\ \printead{e1} \\ \printead
{e2}} 
\address[b]{M. Schweinberger\\ Department of Statistics\\ Penn State
University\\326 Thomas Building \\ University Park, Pennsylvania 16802
\\ USA \\ \printead{e3}}
\end{aug}

\received{\smonth{3} \syear{2009}}
\revised{\smonth{8} \syear{2009}}

%
\begin{abstract}
A model for network panel data is discussed, based
on the assumption that the observed data are discrete observations
of a\break continuous-time Markov process on the space of all directed graphs
on a given node set, in which changes in tie variables
are independent conditional on the current graph.
The~model for tie changes is parametric and designed
for applications to social network analysis, where
the network dynamics can be interpreted as being generated
by choices made by the social actors represented by the
nodes of the graph.
An algorithm for calculating the Maximum Likelihood estimator
is presented, based on data augmentation and stochastic approximation.
An application to an evolving friendship network is given
and a small simulation study is presented which suggests that
for small data sets the Maximum Likelihood estimator is
more efficient than the earlier proposed Method of Moments estimator.
\end{abstract}

%

\end{frontmatter}
%

\section{Introduction}\label{sec1}

Relations between social actors can be studied by methods of social
network analysis [e.g., Wasserman
and Faust (\citeyear{WaFa1994}); Carrington, Scott and Wasserman (\citeyear{CaScWa2005})]. Examples
are friendship between pupils in a
school class or alliances between firms. A basic data structure for
social networks is the
directed graph or digraph, where the actors are represented by the
nodes, and the arcs between the
nodes indicate the social ties. Social network analysis traditionally
has had a focus on rich
description of network data, but the recent development of methods of
statistical inference for
network data [e.g., Airoldi et al. (\citeyear{AiBlFiGoXiZh2007}); Hunter and Handcock
(\citeyear{HuHa2006})]
has the potential of moving this field toward a wider use of
inferential approaches. Longitudinal studies are especially important
for obtaining insight into
social networks, but statistical methods for longitudinal network data
that are versatile enough for
realistic modeling are only just beginning to be developed.

This article considers repeated observations of a relation,
or network, on a
given set of actors $\mathcal{N} = \{1, \ldots, n\}$, observed
according to a panel design,
and represented as a sequence of digraphs
$x(t_m)$ for $m = 1, \ldots, M$, where $t_1 < \cdots< t_M$ are the
observation moments.
The~nodes represent social actors (which may be
individuals, companies, etc.),
and the node set $\mathcal{N}$ is the same for all
observation moments.
A digraph is defined here as an irreflexive relation,
that is, a subset $x$ of $\{(i,j) \in\mathcal{N}^2 \mid i \neq j \}$,
and when $(i,j) \in x$ we shall say that there is an arc,
or a tie, from $i$ to $j$.
It often is realistic to assume that
the social network has been developing between the observation moments,
which leads to the assumption that the observations $x(t_m)$ are
realizations of stochastic digraphs $X(t_m)$ embedded in a
continuous-time stochastic process $X(t)$, $t_1 \leq t \leq t_M$.
Holland and Leinhardt (\citeyear{HoLe1977}) proposed to use
continuous-time Markov chains, defined
on the space of all digraphs with a given node set, for modeling
social network dynamics even if the observations are made
at a few discrete time points and not continuously.
Continuous-time Markov chains provide a natural starting point
for modeling longitudinal network data.
Wasserman (\citeyear{Wa1979}, \citeyear{Wa1980}) and
Leenders (\citeyear{Le1995}) elaborated \textit{dyad-independent models}, where the
ties between pairs of actors (\textit{dyads}) develop according to
processes that are mutually independent between dyads. This is not
realistic for social processes, because dependence between the set
of ties among three or more actors can be very strong, as was found
already by Davis (\citeyear{Da1970}) who showed that a basic feature of
many social networks is the tendency toward
transitivity (``friends of my friends are my friends''). Snijders
and van Duijn (\citeyear{SnDu1997}) and Snijders (\citeyear{Sn2001}) proposed so-called
\textit{actor-oriented models}, explained in the next section, which
do allow such higher-order dependencies.

The~actor-oriented models are too complicated for
the calculation of likelihoods or estimators in closed form,
but they represent stochastic processes which can be easily simulated.
This was exploited by the estimation method proposed
in the papers mentioned, which is
a Method of Moments (\textit{MoM}) estimator implemented algorithmically
by stochastic approximation [Robbins and Monro (\citeyear{RoMo1951});
also see, e.g., Kushner and Yin (\citeyear{KuYi2003})].
This estimator usually performs well
[some empirical applications are given in
de Federico (\citeyear{Fe2003}), van de Bunt, van Duijn and Snijders (\citeyear{BuDuSn1999}),
and van Duijn et al. (\citeyear{DuZeHuStWa2003})].

It is to be expected, however, that the statistical efficiency
of the Maximum Likelihood (\textit{ML}) estimator will be greater.
ML estimation also paves the way for
likelihood-based model selection, which will be a marked
improvement on existing methods of model selection.
This article presents an MCMC algorithm for approximating the ML estimator,
combining and extending ideas in Gu and Kong (\citeyear{GuKo1998}), Snijders (\citeyear{Sn2001}),
and Koskinen and Snijders (\citeyear{KoSn2007}),
the latter of which proposed for this model a MCMC algorithm
for Bayesian inference.
Section \ref{sec2} presents the model definition. The~algorithm for
obtaining the ML estimator is described in Section \ref{sec3}.
Section \ref{sec4} reports results of an empirical example
and Section \ref{sec5} presents a very small simulation study
comparing the ML and MoM estimators.
The~paper finishes with an algorithm for
approximating the likelihood ratio test in Section \ref{sec6} and
a discussion in Section \ref{sec7}.

\section{Model definition}\label{sec2}

We assume that repeated observations $x(t_1), \ldots,\break x(t_M)$
on the network are available for some $M \geq2$.
The~network, or digraph, $x$ will be identified with its $n \times n$
adjacency matrix,
of which the elements
denote whether there is a tie from node $i$ to node $j$
($x_{ij} = 1 $) or not ($x_{ij} = 0 $).
Self-ties are not allowed, so that the diagonal is structurally zero.
Random variables are denoted by capitals.
The~stochastic process $X(t)$, in which the observations
$x(t_m)$ are embedded,
is modeled as being right-continuous.

Various models have been proposed,
most of them being Markov processes of some kind.
We focus on actor-oriented models [Snijders (\citeyear{Sn2001})].
Tie-oriented models [Snijders (\citeyear{Sn2006})] can be treated similarly.
The~basic idea of actor-oriented models [Snijders (\citeyear{Sn2001})] is that
the nodes of the graph represent social actors,
who have control,
albeit under constraints, of their outgoing ties;
and the graph develops as a continuous-time Markov process
(even though it is observed only at $M$ discrete time points).
The~constraints are that ties may change only one by one,
and actors do not coordinate their changes of ties.
Thus, at any given moment, one actor $i$ may create one new tie
or delete one existing tie,
where the probability distribution
of such changes depends on the current digraph;
excluded are simultaneous changes such as swapping one tie for another,
or bargaining between actors over ties.
This constraint was proposed already by Holland and Leinhardt (\citeyear{HoLe1977}),
and it has the virtue of splitting up the change process
in its smallest possible constituents.

The~actor-oriented model is further specified as follows;
further discussion and motivation is given in Snijders (\citeyear{Sn2001}).

\begin{enumerate}
\item[1.] \textit{Opportunities for change}
\end{enumerate}

Each actor $i$
gets at stochastically determined moments the opportunity
to change one of the outgoing tie variables $X_{ij}(t)$
($j \in\mathcal{N}$, $j \neq i$).
Since the process is assumed to be Markovian, waiting times
between opportunities have exponential distributions.
Each of the actors $i$ has a \textit{rate function}
$\lambda_i(\alpha,x)$ which defines how quickly this actor
gets an opportunity to change a tie variable, when the current
value of the digraph is $x$, and where
$\alpha$ is a parameter.
At any time point $t$ with $X(t) = x$, the waiting time until the
next opportunity for change by any actor is exponentially distributed
with parameter
%
\begin{equation}
\lambda(\alpha, x) = \sum_i \lambda_i(\alpha,x)  . \label{lambda1}
\end{equation}
Given that an opportunity for change occurs, the probability that it is
actor $i$ who gets the opportunity is given by
%
\begin{equation}
\pi_i(\alpha,x) = \lambda_i(\alpha,x) / \lambda(\alpha,x)  . \label{wi}
\end{equation}
The~rate functions can be constant between observation moments,
or they can depend on functions $r_{ik}(x)$
which may be covariates or positional characteristics
of the actors such as outdegrees $\sum_j x_{ij}$.
A convenient assumption is to use an exponential link function,
%
\begin{equation}
\lambda_i(\alpha,x) = \exp\biggl(\sum_k \alpha_k r_{ik}(x)\biggr)  . \label{lambda2}
\end{equation}

\begin{enumerate}
\item[2.] \textit{Options for change}
\end{enumerate}

When actor $i$ gets the opportunity to make a change,
this actor
has a permitted set $\mathcal{A}_i(x^0)$ of values to which the digraph
may be changed, where $x^0$ is the current value of the digraph.
The~assumption that the actor controls his or her outgoing ties,
but can change only one tie variable at the time,
is equivalent to
\begin{subequation}
\label{Ai}
\begin{equation}
\mathcal{A}_i(x^0)   \subset  \{ x^0 \} \cup\mathcal{A}_i^{\mathrm{r}}(x^0),
\end{equation}
where $\mathcal{A}_i^{\mathrm{r}}(x^0)$ is the set of adjacency matrices
differing from $x$ in exactly one element,
\begin{eqnarray}
\mathcal{A}_i^{\mathrm{r}}(x^0)   &=&   \{x \mid x_{ij} = 1-x^0_{ij} \mbox
{ for one } j \neq i,\nonumber\\[-8pt]\\[-8pt]
&&\hspace*{14pt}\mbox{ and } x_{hk} = x^0_{hk} \mbox{ for all other } (h,k) \} .\nonumber
\end{eqnarray}
\end{subequation}
Including $x^0$ in $\mathcal{A}_i(x^0)$ can be important for
expressing the property that actors who are satisfied with the
current network will prefer to keep it unchanged.
Therefore, the usual model is $\mathcal{A}_i(x^0) = \{ x^0 \} \cup
\mathcal{A}_i^{\mathrm{r}}(x^0)$.
This does not lead to identifiability problems because
the ratio between not making a change and making a change
is not a free parameter, but
fixed by assumption (\ref{choiceprob}) for the conditional
probabilities of the new state of the network.

Some alternatives are models
with structurally impossible ties,
where the impossible digraphs are left out of $\mathcal{A}_i(x^0)$, and
models where the actor is required to make a change
whenever there is the opportunity,
obtained by leaving the current element $x^0$ out of $\mathcal{A}_i(x^0)$.

It is assumed that the network dynamics is driven by
a so-called \textit{objective function}
$f_i(\beta, x^0,x)$ that can be interpreted as the relative attractiveness
for actor $i$ of moving from the network represented by $x^0$
to the network $x$, and where $\beta$ is a parameter.
Under the condition that the current digraph is $x^0$
and actor $i$ gets the opportunity to make a change,
the conditional probability that the next digraph is~$x$~is modeled as
%
\begin{equation}\label{choiceprob}
\hspace*{8pt} p_i(\beta,x^0, x) = \cases{\displaystyle
\exp(f_i(\beta,x^0,x)) /
\sum_{\tilde x} \exp(f_i(\beta,x^0,\tilde x)),&\quad   $x \in\mathcal{A}_i(x^0)$,
\vspace*{2pt}\cr
0, & \quad  $x \notin\mathcal{A}_i(x^0)$,}\hspace*{-8pt}
\end{equation}
where the summation extends over $\tilde x \in\mathcal{A}_i(x^0)$.
This formula can be motivated
by a random utility argument as used in econometrics
[see, e.g., Maddala (\citeyear{Ma1983})], where it is assumed that the actor
maximizes $f_i(\beta, x^0,x)$ plus a random disturbance having a
standard Gumbel distribution.
Assumption (\ref{Ai}) implies that instead of
$p_i(\beta,x^0, x)$ we can also write $p_{ij}(\beta,x^0)$
where the correspondence between $x$ and $j$ is defined as follows:
if $x \neq x^0$, $j$ is the unique element of $\mathcal{N}$
for which $x_{ij} \neq x^0_{ij}$; if $x = x^0$,
$j=i$. This less redundant notation will be used in the sequel.
Thus,
for $j \neq i$,
$p_{ij}(\beta,x^0)$ is the probability that,
under the condition that actor~$i$ has the opportunity to make a change
and the current digraph is $x^0$,
the change will be to $x_{ij} = 1 - x^0_{ij}$ with the rest unchanged,
while $p_{ii}(\beta,x^0)$ is the probability that,
under the same condition, the digraph will not be changed.

The~most usual models are based on objective functions
that depend on $x$ only. This
has the interpretation that actors
wish to maximize a function $f_i(\beta, x)$ independently of ``where
they come from.''
The~greater generality of (\ref{choiceprob}),
where the objective function can depend
also on the previous state $x^0$, makes it possible to model path-dependencies,
or hysteresis,
where the loss suffered from withdrawing a given tie differs from the gain
from creating this tie, even if the rest of the network
has remained unchanged.

Various ingredients for specifying the objective function
were proposed in Snijders (\citeyear{Sn2001}). A linear form is convenient,
%
\begin{equation}
f_i(\beta, x^0,x) = \sum_{k=1}^L \beta_k   s_{ik}(x^0,x)  , \label{fi}
\end{equation}
where the functions
$s_{ik}(x^0,x)$ are determined by subject-matter knowledge
and available social scientific theory.
These functions can represent essential aspects
of the network structure,
assessed from the point of view of actor $i$, such as
%
\begin{eqnarray}
s_{ik}(x^0,x)  &=& \sum_j x_{ij}  \hspace*{97pt}\mbox{(outdegree)}\\
&& \sum_j x_{ij} x_{ji} \hspace*{83pt} \mbox{(reciprocated ties)} \\
&& \sum_{j,k} x_{ij} x_{jk}  x_{ik} \hspace*{69pt} \mbox{(transitive triplets)}\\
&& \sum_j (1-x_{ij}) \max_k x_{ik} x_{kj} \qquad  \mbox{(indirect ties)}\\
&& \sum_j x^0_{ij} x^0_{ji} x_{ij} x_{ji}\hspace*{57pt}  \mbox{(persistent reciprocity);}
\end{eqnarray}
they can also depend on covariates---such as resources
and preferences of the actors, or costs of exchange
between pairs of actors---or combinations
of network structure and covariates.
For example, de Federico (\citeyear{Fe2003}) found in a study of friendship
between foreign exchange students that friendship formation tends to be
reciprocal, with a negative parameter for forming indirect ties
(thus leading to relatively closed networks),
and that friendships are more likely
to be formed
between individuals from the same region
(a covariate effect), but that reciprocation adds less
to the tendency to form ties between
persons from the same region
than between arbitrary individuals
(negative interaction between covariate
and reciprocity).

\begin{enumerate}
  \item[3.] \textit{Intensity matrix; time-homogeneity}
\end{enumerate}

The~model description given above defines $X(t)$ as
a continuous-time Markov process
with $Q$-matrix or intensity matrix [e.g., Norris (\citeyear{No1997})]
for $x \neq x^0$ given by
%
\begin{equation}\label{q}
q(x^0, x) = \cases{
\lambda_i(\alpha, x^0)  p_i(\beta,x^0, x), & \quad if $x \in \mathcal{A}_i(x^0), i \in\mathcal{N}$,\vspace*{2pt}
\cr
0, &\quad  if $x \notin \displaystyle\bigcup_i \mathcal{A}_i(x^0) $.}
\end{equation}
%

The~assumptions do not imply that the distribution of $X(t)$ is stationary.
The~intensity matrix is time-homogeneous, however,
except for time dependence reflected by time-varying components
in the functions $s_{ik}(x^0,x)$.

For the data-collection designs to which this paper is devoted,
where observations are done
at discrete time points $t_1, \ldots, t_M$, these time points can be used
for marking time-heterogeneity of the transition distribution
(cf. the example in Section~\ref{sec6}).
For example, covariates may be included
with values allowed to change at the observation moments.
A special role is played here by the time durations $t_m - t_{m-1}$
between successive observations.
Standard theory for continuous-time Markov chains [e.g., Norris (\citeyear{No1997})]
shows that the matrix of transition probabilities from $X(t_m)$ to $X(t_{m-1})$
is $e^{(t_m - t_{m-1})Q}$, where $Q$ is the matrix with elements~(\ref{q}). Thus, changing the duration $t_m - t_{m-1}$ can be
compensated by multiplication of the rate function $ \lambda_i(\alpha
, x)$
by a constant.
Since the connection between an externally defined real-valued time
variable $t_m$
and the rapidity of network change is tenuous at best,
it usually is advisable [e.g., see Snijders (\citeyear{Sn2001})]
to include a multiplicative
parameter in the rate function which is constant between observation
moments $t_m$
but differs freely between periods $(t_{m-1}, t_m)$.
With the inclusion of such a parameter, the numerical values $t_m$ become
unimportant because changes in their values can be compensated by the
multiplicative rate parameters.

\subsection{Comparison with discrete-time Markov chain models}\label{sec2.1}

Popular models for analyzing cross-sectional network data
are exponential families of graph distributions such as
the Markov model of Frank and Strauss (\citeyear{FrSt1986}), generalized
to the $p^*$ Exponential Random Graph (ERG) model
by Frank (\citeyear{Fr1991}) and Wasserman and Pattison (\citeyear{WaPa1996}),
and elaborated and further specified by
Hunter and Handcock (\citeyear{HuHa2006}) and Snijders et al. (\citeyear{SnPaRoHa2006}).
Discrete-time Markov chain models, as longitudinal models of this kind,
were proposed by Robins and Pattison (\citeyear{RoPa2001}),
Krackhardt and Handcock (\citeyear{KrHa2007}), and Hanneke and Xing (\citeyear{HaXi2007}).
There are a number of essential differences between these models
and the actor-oriented model treated here, with respect
to interpretation as well as statistical procedures.

When applying the actor-oriented model to a sequence of two or more
repeated observations, these observations are embedded in a
continuous-time model.
This has clear face validity in applications where changes in the network
take place at arbitrary moments between observations.
Singer (\citeyear{Si2008}) gives an overview of the use of this principle
for continuously distributed data.
This approach yields a major advantage over the cited
longitudinal ERG models: our probability model
is defined independently of the observational design.
Data from irregularly spaced observation intervals can be analyzed
without the need to make any adaptations.
Another advantage is that the dynamic process is defined
parsimoniously as a function of its
elementary constituents---in this case, the conditional probability
of a single tie change.
The~random utility interpretation of (\ref{choiceprob}), discussed above,
gives an interpretation in terms of myopic optimization
of an objective function to which a random disturbance is added, and
which can be used to represent a ``social mechanism''
that could have given rise to the observed dynamics.

The~discrete-time ERG models constitute exponential families, which has
the advantage
that many standard techniques can be directly applied.
The~distribution of the
continuous-time process $\{X(t),  t_1 \leq t \leq t_M\}$
for the actor-oriented model constitutes
an exponential family, so for discrete observation moments
our model can be regarded
as an incompletely observed exponential family.
For both types of model, inference is computer-intensive and time-consuming
because of the need to implement elaborate MCMC procedures.
The~definition of the model given above can be used
directly to simulate data from the probability distribution, conditional
on an initial state of the network.
This contrasts with models in the ERG family, which can be simulated
only indirectly, by regarding them as the stationary
distribution of an auxiliary Markov chain,
and applying a Gibbs or Metropolis--Hastings algorithm.
The~near degeneracy problem [Snijders et al. (\citeyear{Sn2006})] which
plagues some specifications of ERG models is, in practice,
not a problem for the actor-oriented model.

\section{ML estimation}\label{sec3}

This section presents an algorithm for MCMC
approximation of the maximum likelihood estimate.
Estimation is done
conditional on the first observation $x(t_1)$.
This has the advantage that no model assumptions need to be invoked
concerning the probability distribution that may have led to
the first observed network $x(t_1)$, and the estimated parameters
refer exclusively to the dynamics of the network.

The~algorithm proposed below can be sketched as follows.
For each $m$\break  ($m = 2, \ldots, M$) the observed data are augmented with
random draws from the
sequence of intermediate digraphs $x(t)$ that could have
led from one observation, $x(t_{m-1})$, to the next, $x(t_m)$
(Section~\ref{S_aug}).
These draws can be simulated using a Metropolis--Hastings algorithm
(Section~\ref{S_sim}).
They are then used in the updating steps of a
Robbins--Monro algorithm,
following ideas of Gu and Kong (\citeyear{GuKo1998}),
to find the solution of the likelihood equation (Section~\ref{S_appr}).
These elements are put together in Section~\ref{S_multi}.

\subsection{Augmented data}
\label{S_aug}

The~likelihood for the observed data $x(t_2), \ldots,\break x(t_M)$
conditional on $x(t_1)$ cannot generally
be expressed in a computable form.
Therefore, the observed data will be augmented with data such that
an easily computable likelihood is obtained,
employing the general data augmentation principle
proposed by Tanner and Wong (\citeyear{TaWo1987}).
The~data augmentation can be done for each period
$(t_{m-1}, t_m)$ separately and, therefore, this section
considers only the observations $x(t_1)$ and $x(t_2)$.

Denote the time points of an opportunity for change
by $T_r$ and their total number between
$t_1$ and $t_2$ by $R$,
the time points being ordered increasingly so that
$t_1 = T_0 < T_1 < T_2 < \cdots< T_R < t_2$.
The~model assumptions imply that at each time~$T_r$,
there is one actor, denoted $I_r$,
who gets an opportunity for change
at this time moment. Define $J_r$
as the actor toward whom the tie variable is changed,
and define formally $J_r = I_r$ if there is no change
[i.e., if $x(T_r) = x(T_{r-1}$)]:
\[
\begin{tabular}{@{}p{245pt}@{}}
$(I_r, J_r)$   is the only $(i,j)$  for which
$x_{ij}(T_{r-1}) \neq x_{ij}(T_r)$ if there is such an $(i,j)$;   else $J_r = I_r$.
\end{tabular}
\]
Given $x(t_1)$, the outcome of the stochastic process
$(T_r, I_r, J_r)$, $r = 1, \ldots, R$
completely determines $x(t),   t_1 < t \leq t_2$.

The~augmenting data consist of $R$ and
$(I_r, J_r)$, $r = 1, \ldots, R$, without the time points $T_r$.
It may be noted that this differs from the definition
of a sample path in Koskinen and Snijders (\citeyear{KoSn2007})
in that the times in between opportunities for change are
integrated out; the reason is to obtain a simpler MCMC algorithm.
The~possible outcomes of the augmenting data are determined by the condition
%
\begin{eqnarray}\label{x2}
\sharp\{r \mid1 \leq r \leq R, (i_r, j_r) = (i,j) \} =\cases{
\mbox{even,}&\quad if $x_{ij}(t_2) = x_{ij}(t_1)$, \vspace*{2pt}\cr
\mbox{odd,}&\quad if $x_{ij}(t_2) \neq x_{ij}(t_1)$,}
\end{eqnarray}
for all $(i,j)$ with $i \neq j$.
The~stochastic process $V = ((I_r, J_r)$, $r = 1, \ldots, R )$
will be referred to as the sample path;
the elements for which $I_r = J_r$, although redundant
to calculate $x(t_2)$ from $x(t_1)$,
are retained because they facilitate the computation of the likelihood.
Define $x^{(k)} = x(T_k)$; the digraphs
$x^{(k)}$ and $x^{(k-1)}$ differ in element $(I_k, J_k)$
provided $I_k \neq J_k$, and in no other elements.

The~probability function of the sample path,
conditional on $x(t_1)$, is given by
%
\begin{eqnarray}\label{full_like1}
&&p_{\mathrm{sp}}\{ V = ((i_1,j_1), \ldots, (i_R,j_R)
) ; \alpha, \beta\} \nonumber\\
& &\qquad  =P_{\alpha, \beta}\bigl\{T_{R} \leq t_2 < T_{R+1} \vert x^{(0)}, (i_1,
j_1), \ldots, (i_R,j_R) \bigr\} \\
&&\quad \qquad {} \times  \prod_{r=1}^{R} \pi_{i_r}\bigl(\alpha, x^{(r-1)}\bigr)  p_{i_r,
j_r}\bigl(\beta, x^{(r-1)}\bigr)  ,\nonumber
\end{eqnarray}
where $\pi_i$ is defined in (\ref{wi}) and
$p_{ij}$ in and just after (\ref{choiceprob}).
Denote the first component of (\ref{full_like1}) by
\begin{eqnarray}\label{kappa1}
&&\kappa\bigl(\alpha, x^{(0)}, (i_1, j_1), \ldots, (i_R,j_R) \bigr) \nonumber\\[-8pt]\\[-8pt]
&&\qquad = P_{\alpha, \beta}\bigl\{T_{R} \leq t_2 < T_{R+1} \vert x^{(0)}, (i_1,
j_1), \ldots, (i_R,j_R) \bigr\}  .\nonumber
\end{eqnarray}
Conditioning on $ x^{(0)}, (i_1, j_1), (i_2, j_2), \ldots, $
[and not on $x(t_2)$!],
the differences $T_{r+1} - T_{r}$
are independently exponentially distributed
with parameters\break $\lambda(\alpha, x^{(r)})$.
Hence, under this conditioning the distribution of $T_R - t_1$ is the
convolution of exponential
distributions with parameters $\lambda(\alpha, x^{(r)})$
for $r = 0, \ldots, R-1$.
In the special case that the actor-level
rates of change $\lambda_i(\alpha,x)$ are constant, denoted by~$\alpha_1$,
$R$~has a Poisson distribution with parameter $n \alpha_1(t_2-t_1)$;
(\ref{kappa1}) then is given by
%
\begin{equation}
\hspace*{6pt}\kappa\bigl(\alpha, x^{(0)},(i_1, j_1), \ldots, (i_R,j_R) \bigr) =
\exp\bigl(-n \alpha_1 (t_2 - t_1) \bigr) \frac{(n \alpha_1
(t_2-t_1))^R}{R!}.\hspace*{-6pt}
\label{kappa4}
\end{equation}
In the general case where the change rates are nonconstant,
the probability (\ref{kappa1}) can be approximated as follows.
Denote by $p_{T_{R}}(t)$ the density function of $T_{R}$,
conditional on $ x^{(0)}, (i_1, j_1)$, $(i_2, j_2), \ldots, $
or, equivalently, on
the embedded process $ x^{(0)}, x^{(1)}, x^{(2)}, \ldots.$
The~distribution of $T_R - t_1$ is a convolution of exponential
distributions and, therefore, the central limit theorem implies that
the density function $p_{T_{R}}(t)$ is approximately the normal density
with expected value
%
\begin{equation}
\mu_\alpha= \sum_{r=0}^{R-1} \bigl\{\lambda\bigl(\alpha, x^{(r)}\bigr)\bigr\}^{-1}
\end{equation}
and variance
%
\begin{equation}
\sigma^2_\alpha= \sum_{r=0}^{R-1} \bigl\{\lambda\bigl(\alpha,
x^{(r)}\bigr)\bigr\}^{-2}  .
\end{equation}
Hence,
%
\begin{equation}
p_{T_{R}}(t)   \approx  \frac{1}{\sqrt{(2\pi\sigma^2_\alpha)}}
  \exp\biggl( \frac{-(t - t_1 - \mu_\alpha)^2}{2\sigma^2_\alpha
}\biggr)  .
\end{equation}
Probability (\ref{kappa1}) now can be expressed as
%
\begin{eqnarray}\label{kappa2}
&& \kappa\bigl(\alpha, x^{(0)}, (i_1, j_1), \ldots, (i_R,j_R) \bigr)
\nonumber\\
&&\qquad = \int_{t_1}^{t_2} p_{T_{R}}(s)   P\{T_{R+1} - T_{R} > t_2 - T_{R}
\mid T_{R} = s \} \, ds \nonumber\\
&&\qquad = \int_{t_1}^{t_2} p_{T_{R}}(s)   \exp\bigl(-\lambda\bigl(\alpha,
x^{(R)}\bigr) (t_2 - s)\bigr)  \, ds \\
&&\qquad  \approx p_{T_{R}}(t_2)   \int_{t_1}^{t_2} \exp\bigl(-\lambda
\bigl(\alpha, x^{(R)}\bigr) (t_2 - s)\bigr) \, ds \nonumber\\
&&\qquad  \approx \frac{ p_{T_{R}}(t_2)}{\lambda(\alpha, x^{(R)})}. \nonumber
\end{eqnarray}
The~approximations are valid for large $R$ and $t_2 - t_1$,
under boundedness conditions on the rate functions $\lambda_i$.
The~first approximation in (\ref{kappa2}) is based on splitting the integration
interval into two intervals $(t_1, t_2 - L)$ and $(t_2 - L, t_2)$
for a bounded but large $L$; the first interval then gives an
asymptotically negligible contribution and on the second interval
$p_{T_{R}}(s)$ is approximately constant
since $\operatorname{var}(T_R) = \mathcal{O}(R)$.
The~second approximation uses that $t_2 - t_1$ is large.
Combining the preceding equations yields
\begin{eqnarray}\label{kappa3}
&&\kappa(\alpha, x^{(0)}, (i_1, j_1), \ldots, (i_R,j_R) ) \nonumber\\[-8pt]\\[-8pt]
&&\qquad \approx \frac{1}{\lambda(\alpha, x^{(R)}) \sqrt{(2\pi\sigma
^2_\alpha)}}
  \exp\biggl( \frac{-(t_2 - t_1 - \mu_\alpha)^2}{2\sigma^2_\alpha
}\biggr)  .\nonumber
\end{eqnarray}
%

This shows that, for observed data $(x(t_1), x(t_2))$ augmented
by the sample path, the likelihood conditional on $x(t_1)$ can be
expressed directly,
either exactly or in a good approximation.

\subsection{Missing data principle and stochastic approximation}
\label{S_appr}

An MCMC algorithm will be used that finds the ML estimator
based on augmented data.
Several methods for MCMC maximum likelihood estimation have been proposed
in the literature.
We shall use
the Markov Chain Stochastic Approximation
(MCSA) algorithm proposed by Gu and Kong (\citeyear{GuKo1998}) and used
(in a slightly different specification) also by Gu and Zhu (\citeyear{GuZh2001}).
This algorithm is based on the missing information principle of
Orchard and Woodbury (\citeyear{OrWo1972}) and Louis (\citeyear{Lo1982})---going back to Fisher (\citeyear{Fi1925}); cf. Efron (\citeyear{Ef1977}).
The~principle can be summarized as follows.
Suppose that $x$ is observed,
with probability distribution parameterized by $\theta$ and having
probability density
$p_X(x; \theta)$ w.r.t. some $\sigma$-finite measure.
To facilitate estimation, the observed data
is augmented by extra data $v$ (regarded as missing data)
such that the joint density is $p_{XV}(x,v; \theta)$.
Denote the observed data score function
$\partial\log(p_{X}(x; \theta)) / \partial\theta$
by $S_X(\theta; x)$ and the total data score function
$\partial\log(p_{XV}(x,v; \theta)) / \partial\theta$
by $S_{XV}(\theta; x, v)$.
It is not hard to prove (see the cited literature) that
%
\begin{equation}
E_\theta\{S_{XV}(\theta; x, V) \mid X=x \} = S_{X}(\theta; x)   .
\label{MI1}
\end{equation}
This is the first part of the missing information principle.
This equation implies that the likelihood equation
can be expressed as
%
\begin{equation}
E_\theta\{S_{XV}(\theta; x, V) \mid X=x \} = 0 , \label{MLE1}
\end{equation}
and, therefore, ML estimates can be determined,
under regularity conditions,
as the solution of (\ref{MLE1}).

This is applied in situations where the observed data
score function $S_X(\theta; x)$ is too difficult to calculate,
while the total data score function $S_{XV}(\theta; x, v)$ is computable.
In our case, we condition on $X(t_1)$, so this is treated
as being fixed; we observe $X = (X(t_2), \ldots, X(t_M))$; and
between each pair of consecutive observations
$X(t_{m-1})$ and $X(t_m)$
the data are augmented, as discussed in the previous section,
by the sample path that could have brought the network
from~$X(t_{m-1})$ to~$X(t_m)$.
These sample paths combined for $m = 2$ to $M$ constitute $V$.
The~following two subsections supply the additional elements
for how the augmenting data are used.

In the MCSA algorithm of Gu and Kong (\citeyear{GuKo1998}), the solution
of (\ref{MLE1}) is obtained by stochastic approximation
[Robbins and Monro (\citeyear{RoMo1951}); Kushner and Yin (\citeyear{KuYi2003})]
which is defined by the updating step
%
\begin{equation}
{\hat{\theta}}^{(N+1)} =
{\hat{\theta}}^{(N)} + a_N D^{-1}
S_{XV}\bigl({\hat{\theta}}^{(N)} ;   x, V^{(N)}\bigr),
\label{RM1}
\end{equation}
where $V^{(N)}$ is generated according to the conditional distribution
of $V$, given \mbox{$X=x$}, with parameter value ${\hat{\theta}}^{(N)}$.
The~sequence $a_N$, called the gain sequence,
consists of positive numbers, tending to 0.
The~matrix $D$ is a suitable matrix.
It is efficient [see Kushner and Yin (\citeyear{KuYi2003})] to use a gain sequence
tending to zero
as $a_N \sim N^{-c}$ for a $c < 1$,
and to estimate $\theta$
not by the last element ${\hat{\theta}}^{(N)}$ produced by the algorithm,
but by a tail average $(N-n_0+1)^{-1} \sum_{n=n_0}^N {\hat{\theta}}^{(n)}$.
For fixed $n_0$ and $N \rightarrow\infty$, this average converges
to the solution of (\ref{MLE1}) for a wide range
of positive definite matrices $D$.

The~second part of the missing information principle
[Orchard and Woodbury (\citeyear{OrWo1972})]
is that the observed Fisher information matrix for the observed data
can be expressed as
\begin{eqnarray}\label{FI2}
D_X(\theta) &=& - \partial S_{X}(\theta;  x) / \partial\theta
\nonumber\\[-8pt]\\[-8pt]
&=& E_\theta  \{D_{XV}(\theta) \mid X=x \} -
\mbox{Cov}_\theta  \{ S_{XV}(\theta;  x,V) \mid X=x \} , \nonumber
\end{eqnarray}
where $D_{XV}(\theta)$ is the complete data observed Fisher
information matrix,
%
\begin{equation}
D_{XV}(\theta) = - \partial S_{XV}(\theta;  x, V) / \partial
\theta  . \label{FI1}
\end{equation}
Expression (\ref{FI2}) can be interpreted loosely as ``information is
total (but partially unobserved) information
minus missing information.''
This formula allows us to calculate standard errors.

\subsection{Simulating the sample path}
\label{S_sim}

The~MCSA method relies on Monte Carlo simulations of
the missing data $V$.
The~Markov property allows us to treat all periods
$(t_{m-1}, t_m)$ separately and, therefore, we simplify the
treatment and notation here by treating only one of those
periods, temporarily assuming that $M=2$.
The~missing data then is
the sample path $((I_1, J_1), \ldots, (I_R, J_R) )$
which specifies the sequence of tie changes that brings the network
from $X(t_{1})$ to $X(t_2)$.
The~advantage of having integrated out the time steps $T_r$---with the use of the expressions
(\ref{kappa4}) and (\ref{kappa2})---is that less random noise is introduced, and
the simulated variable~$V$ is discrete rather than
including a Euclidean vector with
a varying dimension $R$,
which would require a more complicated MCMC procedure.

The~set of all sample paths
connecting $x(t_1)$ and $x(t_2)$ is the set of all
finite sequences
of pairs $(i,j)$, $i,j \in{\mathcal N}$, where
for $i \neq j$ the parity of the number of occurrences
of $(i,j)$ is given by (\ref{x2}).
Such sequences will be denoted by $\underline{v}= (
(i_1,j_1),\ldots, (i_R,j_R))$, and
the set of all these sequences is denoted $\mathcal{V}$.
The~probability of the sample path
conditional on $X(t_1) = x(t_1), X(t_2) = x(t_2)$ is proportional to
(\ref{full_like1}),
rewritten here as
%
\begin{equation}
\kappa\bigl(\alpha, x^{(0)},(i_1, j_1), \ldots, (i_R,j_R) \bigr)
\times  \prod_{r=1}^{R} \pi_{I_r}\bigl(\alpha, x^{(r-1)}\bigr)  p_{I_r,
J_r}\bigl(\beta, x^{(r-1)}\bigr)  ,
\label{full_like2}
\end{equation}
where $\kappa$ is given by (\ref{kappa4}) or
approximated by (\ref{kappa3}),
depending on the specification of $\lambda_i$.
Draws from this distribution can be generated
by the Metropolis--Hastings algorithm, provided that a proposal distribution
is used which connects any two elements of $\mathcal{V}$.

Such a proposal distribution will now be given,
denoting the proposal probabilities by $u(\tilde{\underline{v}}\vert
\underline{v})$ and the
target probabilities, which are proportional to (\ref{full_like2}), by~$p(\underline{v})$.
Then the acceptance probabilities in the Metropolis--Hastings algorithm,
for a current state $\underline{v}$ and a proposed state $\tilde
{\underline{v}}$, are
%
\begin{equation}
\min\biggl\{1, \frac{p(\tilde{\underline{v}})  u(\underline
{v}\vert\tilde{\underline{v}})}{p(\underline{v})  u(\tilde
{\underline{v}}\vert\underline{v})} \biggr\}  . \label{mh}
\end{equation}

A proposal distribution is used that consists of
small changes in $\underline{v}$.
The~construction of the proposal distribution
was based on the considerations that
the proposal distribution should mix well
in the set of all sample paths,
and the Metropolis--Hastings ratios in (\ref{mh})
should be computable relatively easily.
This led to proposal distributions consisting of
the following types of small changes in $\underline{v}$:

\begin{enumerate}
\item \textit{Paired deletions}.
Of all pairs of indices $r_1, r_2$ with
$(i_{r_1}, j_{r_1}) = (i_{r_2}, j_{r_2})$,\break
$i_{r_1} \neq j_{r_1}$,
one pair is randomly selected, and
$(i_{r_1}, j_{r_1})$ and $(i_{r_2}, j_{r_2})$
are deleted from $\underline{v}$.
\item \textit{Paired insertions}.
A pair $(i,j) \in\mathcal{N}^2$ with $i \neq j$
is randomly chosen,
two indices $r_1, r_2$ are randomly chosen,
and the element $(i,j)$ is inserted immediately before $r_1$
and before $r_2$.
\item \textit{Single insertions}.
At a random place in the path (allowing beginning and end),
the element $(i,i)$ is inserted for a random $i \in\mathcal{N}$.
\item \textit{Single deletions}.
Of all elements $(i_r, j_r)$ satisfying $i_r = j_r$,
a randomly chosen one is deleted.
\item \textit{Permutations}.
For randomly chosen $r_1 < r_2$,
where $r_2 - r_1$ is bounded from above by some convenient
number to avoid too lengthy calculations,
the segment of consecutive elements $(i_r, j_r), r = r_1, \ldots, r_2$
is randomly permuted.
\end{enumerate}
It is evident that these five operations yield new sequences within the
permitted set $\mathcal{V}$ [cf. (\ref{x2})].
Paired insertions and paired deletions are each others' inverse operations,
and the same holds for single insertions and single deletions.
These four types of operation together are sufficient for all elements
of $\mathcal{V}$
to communicate.
Permutations are added to achieve better mixing properties.

The~detailed specification of how these elements are
combined in the proposal distribution
can be obtained from the authors.
Two considerations guided this specification.
In the first place, transparency of the algorithm.
Paired insertions and paired deletions are not always unique inverses
of each other. Nonuniqueness of the inverse operation
would lead to complicated counting procedures
to determine proposal probabilities. Therefore, the restriction is made
that pairs of elements $(i,j)$ can only be inserted at, or deleted
from a pair of positions in the sequence,
if there are no occurrences of the same $(i,j)$ between these positions;
and only if at least one other $(i',j')$ (with $i' \neq i$ or $j' \neq j$)
occurs in between these positions.
In this restricted set of operations, paired insertions and paired
deletions are each other's
unique inverses, which simplifies proposal probabilities.
Due to the presence of permutations, all elements of the space
$\mathcal{V}$ still
are reachable from any element.

The~second consideration is computational efficiency.
When proposing that a pair of elements $(i,j)$ is inserted or deleted
at certain
positions, then also proposing to permute a segment between those positions
entails no increase in computational load
for calculating the Metropolis--Hastings ratios, and this permutation
will lead to larger changes in $\underline{v}$,
and thereby, hopefully, better mixing for a given number of computations.

\subsection{Putting it together}
\label{S_multi}

This subsection combines the elements presented in the preceding subsections
to define an algorithm for MCMC approximation of the ML estimate.
It is now assumed that an arbitrary number $M \geq2$ of repeated
observations has been
made: $x(t_1), x(t_2), \ldots, x(t_M)$.
As argued before, we condition on the first observation $x(t_1)$.
The~further data is denoted by
$x = (x(t_2), \ldots, x(t_M))$.
The~parameter $(\alpha, \beta)$ is denoted by $\theta$.

The~Markov property entails that
the observed data score function, conditional on $X(t_1) = x(t_1)$,
can be decomposed as
\begin{eqnarray}\label{score_dec}
&&S_{X \mid X(t_1)}  (\theta; x(t_2), \ldots, x(t_M) \vert
x(t_1)) \nonumber\\[-8pt]\\[-8pt]
&&\qquad  =   \sum_{m=2}^M S_{X(t_m) \vert  X(t_{m-1})}(\theta; x(t_m)
\vert  x(t_{m-1}) )  ,
\nonumber
\end{eqnarray}
where $S_{X(t_m) \vert  X(t_{m-1})}$ is the score function based on the
conditional
distribution of $X(t_{m})$, given $X(t_{m-1})$.

For the period from $t_{m-1}$ to $t_m$, the data set is augmented by
$V_m,$ which defines the order in which ties are changed between
these time points, as described in Section~\ref{S_aug};
this can be denoted by
\[
V_m = ((I_{m1}, J_{m1}), \ldots, (I_{mR_m}, J_{mR_m}) )
\]
with outcome $v_m$.
The~augmenting variable
as used in Section~\ref{S_aug} is $V = (V_2, \ldots, V_M)$.
%
Denote the probability (\ref{full_like1})
for the period from $t_{m-1}$ to $t_m$ by $p_m(v_m; \theta\vert
x(t_{m-1}) )$,
and the corresponding total data score function by
%
\begin{equation}
S_m(\theta; x(t_{m-1}), v_m )  =
\frac{\partial\log  p_m(v_m; \theta\vert  x(t_{m-1}) )}
{\partial\theta}  . \label{MLE3}
\end{equation}
From (\ref{MI1}) and (\ref{score_dec}), and using the Markov property,
it can be concluded that
the observed data score function now can be written as
\begin{eqnarray}\label{MLE4}
\quad &&S_{X \mid X(t_1)}(\theta;    x \vert  x(t_1))   \nonumber\\[-8pt]\\[-8pt]
&&\qquad =\sum_{m=2}^M  E_\theta\{ S_m(\theta; x(t_{m-1}), V_m )
\vert  X(t_{m-1}) = x(t_{m-1}), X(t_m) = x(t_m) \}  .
\nonumber\hspace*{-20pt}
\end{eqnarray}
The~ML estimate is obtained as the value of $\theta$ for which (\ref{MLE4})
equals 0 [cf. (\ref{MLE1})].

The~algorithm is iterative,
and the $N$th updating step now can be represented as follows:
\begin{enumerate}
\item For each $m = 2, \ldots, M$, make a large number of the
Metropolis--Hastings steps
as described in Section~\ref{S_sim}, yielding $v^{(N)} =
(v_2^{(N)}, \ldots, v_M^{(N)})$.
\item Compute 
\[
S_{XV}\bigl({\hat{\theta}}^{(N)} ;   x, v^{(N)}\bigr) =
\sum_{m=2}^M S_m\bigl({\hat{\theta}}^{(N)} ; x(t_{m-1}), v_m^{(N)}
\bigr)  ,
\]
using (\ref{MLE3})
with $p_m(v_m; \theta\vert  x(t_{m-1}) )$ defined by
(\ref{full_like1}) for the period from $t_{m-1}$ to $t_m$, and
using (\ref{kappa1}) and (\ref{kappa4}) or (\ref{kappa3}).
\item Update the provisional parameter estimate by
\[
{\hat{\theta}}^{(N+1)} =
{\hat{\theta}}^{(N)} + a_N D^{-1}
S_{XV}\bigl({\hat{\theta}}^{(N)} ;   x, v^{(N)}\bigr)  .
\]
\end{enumerate}
As mentioned in Section \ref{S_appr}, the estimate
${\hat{\theta}}_{\mathrm{ML}}$
is calculated as a
tail average of the values ${\hat{\theta}}^{(N)}$ generated by this algorithm.
The~covariance matrix of the ML estimator
is estimated using (\ref{FI2}), where the
expected values in the right-hand side are approximated by Monte Carlo
simulation
of the conditional distribution of~$V$ given $X=x$,
for $\theta= {\hat{\theta}}_{\mathrm{ML}}$.
This involves the matrix of partial derivatives (\ref{FI1}),
which can be estimated by a
score-function method as elaborated in Schweinberger and Snijders (\citeyear{ScSn2006}).

The~main implementation details are the following:

\begin{enumerate}
\item[a.]
The~Method of Moments (MoM) estimator [Snijders (\citeyear{Sn2001})], in practice, yields
a good initial value ${\hat{\theta}}^{(1)}$.

\item[b.]
To avoid long burn-in times for step (1.), the Metropolis--Hastings algorithm
for generating $V^{(N)}$ can be started from the previous value,
$V^{(N-1)}$, rather than independently.
This leads to some autocorrelation in the updates defined in step~(3.),
and the number of Metropolis--Hastings steps must be large
enough that this autocorrelation is not too high.
It may be noted that the Robbins Monro algorithm is robust to moderate
dependence between successive updates [Kushner and Yin (\citeyear{KuYi2003})].
We have found good results when the number of steps
is tuned so that the autocorrelations between the elements of ${\hat
{\theta}}^{(N)}$
are less than~0.3.

\item[c.]
For $D$ in step (3.) we use a Monte Carlo estimate of
the complete data observed Fisher information matrix (\ref{FI1}),
estimated for $\theta= {\hat{\theta}}^{(1)}$ before making
the iteration steps.

\item[d.]
For the other numerical parameters of the algorithm
we use the same values as described, for the stochastic approximation algorithm
to compute the MoM estimator, in Snijders (\citeyear{Sn2001}).
\end{enumerate}

This algorithm is implemented in Siena version 3.3 [Snijders et al. (\citeyear{SnStScHu2009})]. The~executable program as well as the code can be
found at\break
\url{http://www.stats.ox.ac.uk/siena/}.

\section{Empirical example}
\label{sec4}

As an illustrative example, the data set is used that was also analyzed in
van de Bunt, van Duijn and Snijders (\citeyear{BuDuSn1999}) and in Snijders (\citeyear{Sn2001}).
This is a friendship network between 32 freshman students in a given discipline
at a Dutch university.
Initially most of the students were unknown to each other.
There were six waves denoted $t_1 - t_6$ of data collection,
with 3 weeks between waves for the start of the academic year,
and 6 weeks in between later. The~relation studied
is ``being friends or close friends.''
The~data set is obtainable from website
\href{http://www.stats.ox.ac.uk/siena/}{http://www.stats.ox.ac.uk/}
\href{http://www.stats.ox.ac.uk/siena/}{siena/}.

The~transitions between observations
$t_2$ to $t_3$, and $t_3$ to $t_4$,
will here be studied separately.
To identify the rate function,
we assume (arbitrarily but conveniently)
that the duration of the time periods is unity,
$t_3 - t_2 = t_4 - t_3 = 1$.
To keep the model specification relatively simple,
the rate function is supposed to be constant across actors,
given by $\alpha_1$ from $t_2$ to $t_3$ and
by $\alpha_2$ from $t_3$ to $t_4$;
and the objective function~(\ref{fi}) is chosen independent
of the previous state $x^0$, containing contributions
of the effects of outdegree, number of reciprocated ties,
number of transitive triplets, number of 3-cycles,
gender of the sender of the tie (``ego''), gender of the
receiver of the tie (``alter''), and gender similarity:
\begin{eqnarray*}
f_i(\beta, x) &=& \beta_1   \sum_j x_{ij}
 +  \beta_2  \sum_j x_{ij} x_{ji} \\
&&{}+ \beta_3  \sum_{j,k} x_{ij} x_{jk}  x_{ik} \nonumber
 +  \beta_4  \sum_{j,k} x_{ij}   x_{jk} x_{ki}
 +  \beta_{5}   \sum_j x_{ij}   (z_{j} - \bar z) \\
\nonumber
&&{}+ \beta_{6}   \sum_j x_{ij}   (z_{i} - \bar z)
+ \beta_{7}   \sum_j x_{ij}   \{1 - \vert z_{i} - z_{j} \vert
\}  , \nonumber
\end{eqnarray*}
where variable $z_{i}$ indicates the gender of actor $i$ ($F = 0$, $M = 1$)
and $\bar z$ its average over the 32 individuals.
This model illustrates two types of triadic dependence
(transitive triplets and 3-cycles) and
the use of covariates (gender).

For this model the parameters are estimated for the two
transitions $t_2 - t_3$ and $t_3 - t_4$ separately
both by the Method of Moments (MoM) and by Maximum Likelihood.
For models where the functions $s_{ik}$ used in (\ref{fi})
depend only on $x$ and not on~$x^0$, so that they can be expressed
as $s_{ik}(x)$, the MoM estimator as defined by Snijders (\citeyear{Sn2001}) is
based on the vector
with components $\sum_{i,j} \vert X_{ij}(t_m) - X_{ij}(t_{m-1})\vert$
for $m = 2, \ldots, M$ and
$\sum_{m=2}^M \sum_i s_{ik}(X(t_m))$ for $k = 1, \ldots, L$.
The~MoM estimator is defined as the parameter vector for which the observed
and expected values of this statistic are equal, and can be determined
by stochastic approximation.

Both the moment equation and the likelihood equation
can be represented as $E_\theta S = 0 $,
where $S$ is the difference between observed and estimated moments
or the complete-data score function, respectively.
For both estimators, after running the stochastic approximation
algorithm, convergence was checked by simulating the
model for the obtained parameters for 2000 runs and
calculating, for each component $S_h$ of $S$,
the ratio of the average simulated $S_h$ to
its standard deviation. In all cases, this ratio was less than 0.1,
indicating adequate convergence.

%
%
%
%
%
%
%
%
%
%
%
%
%

\begin{table}[b]
\caption{Parameter estimates (Method of Moments and Maximum Likelihood)
for data set of van de Bunt, van Duijn and Snijders (\protect\citeyear{BuDuSn1999})} \label{tab1}
\begin{tabular*}{\textwidth}{@{\extracolsep{\fill}}lld{2.2}cd{2.2}cd{2.2}cd{2.2}c@{}}
\hline
& \textbf{Effect}& \multicolumn{4}{c}{$\bolds{(t_2, t_3)}$}& \multicolumn{4}{c@{}}{$\bolds{(t_3, t_4)}$} \\[-6pt]
& & \multicolumn{4}{c}{\hrulefill}& \multicolumn{4}{c@{}}{\hrulefill} \\
& & \multicolumn{2}{c}{\textbf{MoM}}& \multicolumn{2}{c}{\textbf{ML}} & \multicolumn{2}{c}{\textbf{MoM}}& \multicolumn{2}{c@{}}{\textbf{ML}} \\[-6pt]
& & \multicolumn{2}{c}{\hrulefill}& \multicolumn{2}{c}{\hrulefill} & \multicolumn{2}{c}{\hrulefill}& \multicolumn{2}{c@{}}{\hrulefill} \\
&  & \multicolumn{1}{c}{\textbf{Est.}} & \multicolumn{1}{c}{\textbf{S.E.}}&\multicolumn{1}{c}{\textbf{Est.}} & \multicolumn{1}{c}{\textbf{S.E.}}
& \multicolumn{1}{c}{\textbf{Est.}} & \multicolumn{1}{c}{\textbf{S.E.}}& \multicolumn{1}{c}{\textbf{Est.}} & \multicolumn{1}{c@{}}{\textbf{S.E.}}\\
\hline
\multicolumn{2}{@{}l}{\textit{Rate function}} &  \\
$\alpha$ & Rate parameter &  3.95 & 0.64 & 2.61 & 0.36 &3.43 & 0.59 & 5.67 & 0.75 \\
\multicolumn{2}{@{}l}{\textit{Objective function}} &  \\
$\beta_1$ & Outdegree & -1.66 & 0.26 & -1.02 & 0.26 & -2.19 & 0.30& -2.00 & 0.22 \\
$\beta_2$ & Reciprocated ties & 2.06 & 0.47 & 1.94 & 0.39 & 2.26 &0.55 & 1.70 & 0.35 \\
$\beta_3$ & Transitive triplets & 0.30 & 0.06 & 0.18 & 0.06 & 0.36 &0.07 & 0.26 & 0.05 \\
$\beta_4$ & 3-cycles & -0.59 & 0.23 & -0.42 & 0.17 & -0.59 & 0.27 &-0.31 & 0.14 \\
$\beta_5$ & Gender alter & 0.28 & 0.30 & 0.14 & 0.36 & 0.70 & 0.39 &0.61 & 0.33 \\
$\beta_6$ & Gender ego & -0.34 & 0.32 & -0.41 & 0.41 & -0.04 & 0.38& -0.10 & 0.33 \\
$\beta_7$ & Gender similarity & 0.33 & 0.30 & 0.34 & 0.36 & 0.48 &0.37 & 0.44 & 0.32 \\
\hline
\end{tabular*}
\end{table}

Parameter estimates and standard errors are reported
in Table \ref{tab1}.
Assuming that the estimators are approximately normally distributed
(which is supported by the simulations reported in the next section,
although we have no proof),
the significance can be tested by referring the ratios
of estimate to standard error to a standard normal distribution.
The~Method of Moments and Maximum Likelihood estimates are different
but lead to the same substantive conclusions.
The~parameters reflecting network structure, $\beta_1$ to $\beta_4$,
give the same picture for both transitions.
The~negative $\hat\beta_1$ indicates that an outgoing tie which is
not reciprocated
and not transitively embedded is not considered attractive;
the positive $\hat\beta_2$ and $\hat\beta_3$ indicate
that there is evidence for tendencies toward
reciprocation and transitivity of friendship choices,
when controlling for all other effects in the model.
For interpreting the 3-cycles effect, note that
closed 3-cycles are structures denying a hierarchically
ordered relation.
The~negative $\hat\beta_4$ indicates a tendency away from closed
3-cycles, which---in conjunction with the positive
transitive triplets parameter---could be interpreted as a tendency toward
a local (i.e., triadic) popularity hierarchy in friendship.
For gender, there is only a close to significant effect
of gender alter for the $t_3$--$t_4$ transition,
suggesting that male students tend to be more popular
as friends, when controlling for the other effects.

From this single data example, of course
no conclusions can be drawn concerning
the relative value of these two estimation methods.

\section{Simulation examples}
\label{sec5}

A comparison between the finite sample behavior of the MoM
and the ML estimators can be based on simulations.
Here we present a small simulation study
as a very limited exploration of the
relative efficiency of the two estimators,
which is expected to be in favor of the ML estimator.
The~limited nature of this simulation study does not
allow generalization, but the study design is
meant to be typical for
applications to friendship networks in rather small groups,
and replicates approximately the empirical study
of the previous section.

The~model is identical to that of the previous section:
there are three repeated observations, 32 actors, one binary covariate
called gender, and the first observed network
as well as the distribution of the covariate are identical to
the van de Bunt data set at observation $t_2$.
Therefore, the observation moments are again referred to as
$t_2, t_3, t_4$.
The~parameter values are rounded figures close
to the estimates obtained in the preceding section.
Networks for times $t_3$ and $t_4$
are generated and parameters estimated
under the assumption that parameters $\beta_k $ are the same
in periods $(t_2, t_3)$ and $(t_3, t_4)$.
The~simulation model has parameters
$\alpha_1 = 2.5$ and $\alpha_2 = 3.5$,
$\beta_1 = -2, \beta_2 = 1, \beta_3 = 0.2,
\beta_4 = 0, \beta_5 = 0.5, \beta_6 = -0.25$, and $\beta_7 = 0.5$.
A total of 1000 data sets were generated and the estimates calculated
by both methods.
Table~\ref{tab2} reports the average estimates, the root mean
squared errors,
the rejection rates for testing the data-generating value of the
parameter as the null hypothesis (estimating type-I error rates), and
the rejection rates for testing that the parameters equal 0
(estimating power), where the tests were two-sided tests
based on the $t$-ratio for the corresponding parameter estimate,
assuming a standard normal reference distribution,
at a nominal significance level of 5\%.

\begin{table}
\tabcolsep=0pt
\caption{Simulation results, 3 waves for 32 actors: average parameter
estimates (``Ave''),
root mean squared errors (``RMSE''), estimated type-I error rates
(``$\alpha$''),
estimated power (``$\beta$'')} \label{tab2}
\begin{tabular*}{\textwidth}{@{\extracolsep{\fill}}lld{2.3}d{1.3}ccd{2.3}d{1.3}cc@{}}
\hline
& \textbf{Effect}& \multicolumn{4}{c}{\textbf{MoM estimator}} & \multicolumn{4}{c@{}}{\textbf{ML estimator}} \\[-6pt]
& & \multicolumn{4}{c}{\hrulefill} & \multicolumn{4}{c@{}}{\hrulefill} \\
&  & \multicolumn{1}{c}{\textbf{Ave}} & \multicolumn{1}{c}{\textbf{RMSE}}&\multicolumn{1}{c}{$\bolds{\alpha}$} & \multicolumn{1}{c}{$\bolds{\beta}$}&
 \multicolumn{1}{c}{\textbf{Ave}} & \multicolumn{1}{c}{\textbf{RMSE}}& \multicolumn{1}{c}{$\bolds{\alpha}$} & \multicolumn{1}{c}{$\bolds{\beta}$} \\
\hline
\multicolumn{2}{@{}l}{\textit{Rate function}} &  \\
$\alpha_1 = 2.5$ & Rate $t_2 - t_3$ & 2.46 & 0.44 & 0.063 &-- & 2.37 & 0.44 & 0.114 & --\\
$\alpha_2 = 3.5$ & Rate $t_3 - t_4$ & 3.47 & 0.55 & 0.045 &-- & 3.39 & 0.54 & 0.099 & --\\
\multicolumn{2}{@{}l}{\textit{Objective function}} &  \\
$\beta_1 = -2.0$ & Outdegree & -2.01 & 0.15 & 0.053 & 1.00 & -1.96 &0.14 & 0.047 & 1.00 \\
$\beta_2 = 1.0 $ & Reciprocation & 1.03 & 0.26 & 0.044 &0.96 & 0.97 & 0.26 & 0.042 & 0.95 \\
$\beta_3 = 0.2 $ & Transitivity & 0.186 & 0.052 & 0.043 &0.94 & 0.180 & 0.054 & 0.064 & 0.93 \\
$\beta_4 = 0.0 $ & 3-cycles & 0.02 & 0.14 & 0.042&--& 0.03 & 0.14 & 0.060 &--\\
$\beta_5 = 0.5 $ & Gender alter & 0.53 & 0.24 & 0.043 & 0.65& 0.51 & 0.25 & 0.046 & 0.57 \\
$\beta_6 = -0.25 $ & Gender ego & -0.28 & 0.26 & 0.062 & 0.21 & -0.28& 0.27 & 0.059 & 0.20 \\
$\beta_7 = 0.5 $ & Gender sim. & 0.52 & 0.24 & 0.031 & 0.64& 0.52 & 0.25 & 0.052 & 0.60 \\
\hline
\end{tabular*}
\end{table}

Out of the 1000 generated data sets, 5 were excluded because they did
not produce
well converging results in the default specification of the algorithm
for one or both estimators.
Table \ref{tab2} shows that the results for the two estimators are
very similar,
and the type-I error rates are close to the nominal value,
except for the inflated type-I rates of the ML estimators
for the two rate parameters.
The~latter is related to the skewed distribution of the rate parameter
estimators.
The~correlations between the estimators are more than 0.93 for all coordinates;
note that correlations are attenuated due to the stochastic nature of
the algorithms.
It can be concluded that for this type of model, characterized by
32 actors and 3 waves with rates of change 2.5 and 3.5, with 7 parameters
in the objective function, MoM and ML estimation yield quite similar results.

To explore data sets with less information,
a similar simulation study was conducted with 20 actors, where the
first network
was induced by the $t_2$ network of~20 of the actors in the van de Bunt
data set,
and the rest of the simulation design differed from the previous study by
including an interaction effect of reciprocity by gender similarity,
represented by the term
\[
\beta_{8}   \sum_j x_{ij}  x_{ji}  \{1 - \vert z_{i} - z_{j}
\vert \}  ,
\]
with parameter $\beta_8 = 0.5$.
This effect is included to achieve a higher correlation of the
parameter estimators, which together with the smaller number
of actors should lead to greater difficulties in estimation.

\begin{table}
\tabcolsep=0pt
\caption{Simulation results, 3 waves for 20 actors: average parameter
estimates (``Ave''), root mean squared errors (``RMSE''), estimated type-I error rates
(``$\alpha$''), estimated power (``$\beta$'')} \label{tab3}
\begin{tabular*}{\textwidth}{@{\extracolsep{\fill}}lld{2.3}d{1.3}ccd{2.3}d{1.3}cc@{}}
\hline
& \textbf{Effect}& \multicolumn{4}{c}{\textbf{MoM estimator}} & \multicolumn{4}{c@{}}{\textbf{ML estimator}} \\[-6pt]
& & \multicolumn{4}{c}{\hrulefill} & \multicolumn{4}{c@{}}{\hrulefill} \\
&  & \multicolumn{1}{c}{\textbf{Ave}} & \multicolumn{1}{c}{\textbf{RMSE}}&\multicolumn{1}{c}{$\bolds{\alpha}$} & \multicolumn{1}{c}{$\bolds{\beta}$}&
 \multicolumn{1}{c}{\textbf{Ave}} & \multicolumn{1}{c}{\textbf{RMSE}}& \multicolumn{1}{c}{$\bolds{\alpha}$} & \multicolumn{1}{c}{$\bolds{\beta}$} \\
\hline
\multicolumn{2}{@{}l}{\textit{Rate function}} & \\
$\alpha_1 = 2.5$ & Rate $t_2 - t_3$ & 2.49 & 0.50 & 0.059 &-- & 2.37 & 0.48 & 0.093 & --\\
$\alpha_2 = 3.5$ & Rate $t_3 - t_4$ & 3.51 & 0.70 & 0.044 &-- & 3.36 & 0.67 & 0.094 & --\\
\multicolumn{2}{@{}l}{\textit{Objective function}} & \\
$\beta_1 = -2.0$ & Outdegree & -2.09 & 0.29 & 0.009 & 0.98 & -2.02 &0.21 & 0.025 & 1.00 \\
$\beta_2 = 1.0 $ & Reciprocation & 1.08 & 0.35 & 0.043 &0.86 & 0.97 & 0.33 & 0.052 & 0.82 \\
$\beta_3 = 0.2 $ & Transitivity & 0.170 & 0.099 & 0.022 &0.51 & 0.158 & 0.096 & 0.059 & 0.52 \\
$\beta_4 = 0.0 $ & 3-cycles & 0.01 & 0.23 & 0.034&--& 0.05 & 0.23 & 0.057 &--\\
$\beta_5 = 0.5 $ & Gender alter & 0.59 & 0.44 & 0.020 & 0.15& 0.60 & 0.38 & 0.048 & 0.43 \\
$\beta_6 = -0.25 $ & Gender ego & -0.35 & 0.50 & 0.024 & 0.00 & -0.34& 0.36 & 0.034 & 0.11 \\
$\beta_7 = 0.5 $ & Gender sim. & 0.61 & 0.50 & 0.013 & 0.12& 0.63 & 0.43 & 0.051 & 0.35 \\
$\beta_8 = 0.5 $ & G. sim. $\times$ rec. & 0.47 & 0.63& 0.039 & 0.10 & 0.50 & 0.60 & 0.043 & 0.14 \\
\hline
\end{tabular*}
\end{table}

The~results are shown in Table \ref{tab3}.
Of the 1000 generated data sets, 20 were excluded from the results
because of questionable convergence of the algorithm.
The~table shows that the ML estimator here performs
clearly better than the MoM estimator.
The~estimated relative efficiency of MoM compared to ML
expressed as ratio of mean squared errors ranges for these
10 parameters from 0.50 to 0.99.
The~correlation between the estimators ranges from 0.71 (gender ego)
to 0.98 (rate period 1).
For all parameters
except $\beta_2$, the estimated power of the test based on the ML
estimator is higher
than that of the MoM-based test,
which can be traced back to the combination of a smaller mean squared error
and a less conservative test.
The~surprisingly low power of the MoM-based test for the gender-ego
effect reflects high standard errors that tend to be obtained for
estimating $\beta_6$.

\section{Likelihood ratio tests}
\label{sec6}

One of the important advantages of a likelihood approach
is the possibility of elaborating model selection procedures.
Here we only explain how to conduct a likelihood ratio test.

A convenient way to estimate the likelihood ratio is by
a simple implementation of the idea of
the path sampling method described by Gelman and Meng (\citeyear{GeMe1998}),
who took this method from the statistical physics
literature where it is known as thermodynamical integration.
For arbitrary $\theta_0$ and $\theta_1$,
defining the function $\theta(t) = t \theta_1 + (1-t) \theta_0$,
with $\dot\theta= \partial\theta/ \partial t = \theta_1 - \theta_0$,
this method is based on the equation
%
\begin{equation}
\log\biggl( \frac{p_X (x; \theta_1)}{p_X (x; \theta_0)} \biggr)
  =   \int_0^1 \dot\theta S_X(x; \theta(t))\,  d t  .
\label{llr}
\end{equation}
This integral can be approximated by replacing the integral
by a finite sum and using (\ref{MI1}).
Applying a simulation procedure which for each $\theta(h/H)$,
$h = 0, \ldots, H$, generates $L \geq1$ draws $V_{h\ell}$ from
the conditional distribution of $V$ given $X=x$ under parameter
$\theta(h/H)$, the log likelihood ratio (\ref{llr})
can be approximated by
%
\begin{equation}
\frac{(\theta_1 - \theta_0)}{L (H+1)} \sum_{\ell=1}^L \sum_{h=0}^H
S_{XV}\bigl(\theta(h/H); x, V_{h\ell}\bigr)  .
\label{path_samp}
\end{equation}
Burn-in time can be considerably reduced when starting the MCMC algorithm
for generating $V_{h1}$ by the end result of the algorithm
used to generate $V_{(h-1), L}$.

This was applied to the van de Bunt data set used in Section~\ref{sec4}
to test the null hypothesis that all parameters $\beta_1, \ldots,
\beta_7$ are
for period $(t_3, t_4)$ the same as for $(t_2, t_3)$ against the alternative
that they are allowed to be different. The~rate parameters $\alpha$
were allowed
to be different under the null as well as the alternative hypothesis.
For calculating (\ref{path_samp}) we used $L = 10$, $H = 1000$.
The~estimated likelihood ratio was 18.7, which for a $\chi^2_7$ distribution
yields $p < 0.01$, leading to rejecting the null hypothesis at conventional
levels of significance.

\section{Discussion}\label{sec7}

This article has presented a model for
longitudinal network data
collected in a panel design and an algorithm for calculating the
ML estimator.
The~model can represent triadic and other complicated dependencies
that are characteristic for social networks.
It is designed for applications in social network analysis,
but related models may be applied for modeling
networks in other sciences, such as biology.
Earlier, a method of moments (MoM) estimator for this model
was proposed by Snijders (\citeyear{Sn2001})
and Bayesian inference methods by Koskinen and Snijders (\citeyear{KoSn2007}).
The~algorithm was constructed using stochastic approximation
according to the approach proposed by Gu and Kong (\citeyear{GuKo1998}),
and employs Monte Carlo simulations of
the unobserved changes between
the panel waves, conditional on the observed data.
The~simulation design used is more efficient than that
used in Koskinen and Snijders (\citeyear{KoSn2007})
because here the waiting times
between unobserved changes are integrated out.

No proof is yet available for the consistency and asymptotic normality of
the ML estimator, which are intuitively plausible
for the situation where $n$ tends to infinity,
$t_1$ and $t_2$ are fixed, and the
parameters are such that the average degree $n^{-1}\sum_{i,j} x_{ij}$
tends to a positive finite limit.
Limited simulation results do support the expectation
that the estimators are asymptotically normal.
The~nonstandard assumptions (lack of independence)
imply, however, that a proof may be expected to be rather complicated.

Our experience in the reported simulations and in
working with empirical data sets is that the algorithm converges well
unless data sets are small given the number of
estimated parameters, but the algorithm is time-consuming
(e.g., each ML estimation for one data set in Table \ref{tab2} took about 35 minutes
on a regular personal computer, while each MoM estimation took about 2 minutes).
Further improvements in the algorithm are desirable
for the practical use of ML estimators in these models.
This is the subject of future work.

Judging from our very limited simulations,
the advantages of ML estimation over MoM estimation
in terms of root mean squared error and power of the
associated tests seem strong for small data sets
and small for medium to large data sets.
Further simulation studies are necessary.
However, even if the statistical efficiency
is similar, likelihood-based estimation
can have additional advantages, for example,
the possibility of extensions to more complicated models
and of elaborating model selection procedures.

Software implementing the procedures in this paper
is available at\break
\url{http://www.stats.ox.ac.uk/siena/}.

\section*{Acknowledgments}
Part of this work was done
while M. Schweinberger was working at the University of Groningen and the
University of Washington.
Part of this work was done
while J. Koskinen was working at the University of Stockholm and the
University of Melbourne.

\printaddresses

\end{document}